\input psfig.sty
\documentclass{aa}
\begin{document}

   \thesaurus{08.16.7}  
   
   \title{HST/FOC observations confirm the presence of a spectral feature in the
optical spectrum of Geminga.\thanks{Based on observations with the NASA/ESA
Hubble Space Telescope,
obtained at the Space Telescope Science Institute, which is operated
by AURA, Inc., under NASA contract NAS 5-26555}}

\author{R.P. Mignani\inst{1}, P. A. Caraveo\inst{2} and G.F. Bignami\inst{3}} 

   \offprints{R.P. Mignani}

   \institute{Max-Plack-Institute f\"ur Extraterrestrische Physik,
   Postfach 1603 Giessenbachstrasse, D85740-Garching, Germany \\
              email: rmignani@xray.mpe.mpg.de
         \and
             Istituto di Fisica Cosmica del CNR, Via Bassini 15, 20133-Milan, Italy  \\
             email: pat@ifctr.mi.cnr.it
         \and 
	Agenzia Spaziale Italiana, Via di Villa Patrizi 13, 00161-Rome, Italy \\
	email: gfb@asirom.rm.asi.it
             }

 \date{Received ; accepted  }

\titlerunning{ HST/FOC obs. of Geminga.}
\maketitle

%$\clubsuit$ 
%Based on observations with the NASA/ESA Hubble Space Telescope,
%obtained at the Space Telescope Science Institute, which is operated
%by AURA, Inc., under NASA contract NAS 5-26555

   \begin{abstract}
New optical and near-UV HST observations of Geminga are presented. When compared with
previous ground-based and HST imaging, the data confirm  and better define the presence
of an emission feature centered at $\sim 6,000 \AA$ and superimposed  on the thermal
continuum best fitting the extreme-UV/soft X-ray data.  This
feature may be interpreted in terms of cyclotron emission  originated from a mixture of
H/He ions in the neutron star's atmosphere. In the case of pure Hydrogen, 
the  feature wavelength would
imply a magnetic field of order $3-5~10^{11}~G$, consistent with the value deduced from
the dynamical parameters of the pulsar. If due to a cyclotron emission, the observation
of this feature  would represent the first case of an {\it in situ}  measurement of the
surface magnetic field of an isolated neutron star.
   
   \end{abstract}

%
%________________________________________________________________

\section{Introduction}

Our understanding of the optical properties of Isolated Neutron Stars (INSs) is limited
by the  faintness of the vast majority of them.  Indeed, only for the $m_{V} \sim
16.6$  Crab pulsar,  acceptable,  medium-resolution optical spectroscopy is available (Nasuti
et  al 1996).  For few more cases (PSR0540-69, PSR0833-45, PSR1509-58, PSR0656+14 and
Geminga), the spectral information just relies on multicolour photometry (Nasuti et al.
1997; Mignani et al. 1998a;  Pavlov et al. 1997; Bignami et al. 1996).  For the rest of
the optical database (PSR0950+08, PSR1929+10  and PSR1055-52), only one-or two-band detections
are available  (Pavlov et al. 1996; Mignani et al. 1997). \\
Very young objects, say up to $\tau \sim  10^{4}$ yrs (Crab, PSR0540-69, PSR0833-45),
are  characterized by flat, synchrotron-like spectra arising  from energetic electron
interactions in their magnetosphere (Caraveo, 1998). For
the Middle-Aged ($\tau \sim 10^{5}~ yrs$) INSs (PSR0656+14, Geminga and PSR1055-52), 
also referred to as "the three musketeers" for their overall similarities (Becker \&
Tr\"umper 1997), the situation is more complex and different emission processes may
become relevant.  For example, the non-thermal magnetospheric emission could have faded
enough, at least in the optical waveband, to render visible the thermal emission from
the hot neutron star surface (Mignani et al. 1998b).  Its temperature, following
standard cooling calculations (see e.g. Nomoto \& Tsuruta 1987), could be in the
$10^{5}- 10^{6} ~ K$ range, in excellent agreement  with  the recent X-ray findings
(e.g.  Becker \& Tr\"umper 1997). It is easy,  then, to predict the IR-optical-UV
fluxes expected along the $\sim E^{2}$  Rayleigh-Jeans slope of the Planck curve best
fitting the X-rays, and to compare predictions with observations, where available.

%__________________________________________________________________

\section{Data overview.}

Among Middle-Aged INSs, Geminga is certainly the most studied (see Bignami \& Caraveo
1996 and references therein). Here we review ten years of optical data which are summarized in Table 1 as well as in Fig.1. On the basis of 3 colours only (Fig.1a),  Bignami et al. (1988)
realized that the Geminga spectrum could not be fit by a simple Planck's law. The
addition of the ground-based I upper limit, obtained with the NTT, and new HST colours
(namely  555W, 342W and the upper limit at 675W) 
over the years (Fig. 1b) led  Bignami et al. (1996)  to propose
an explanation in terms of a spectral feature superimposed to the Rayleigh-Jeans 
continuum of the Planckian  best-fitting the ROSAT/EUVE data.   Since the distance to
Geminga was fixed in $ 157^{+59}_{-34}$ pc by the parallax measurement of Caraveo
et al. (1996), an absolute fit to its surface thermal emission is possible. Using the
temperature $T = 4.5 ~ 10^{5} ~K$ (Halpern \& Rudermann 1993),
obtained with ROSAT data collected in 1991, the best fit to the B-UV flux values yields for Geminga a radius $R$ of about 10 km, while a radius of 15 km is
required if one uses the temperature best fitting  the EUVE data, i.e.  $2.5 ~ 10^{5}
~K$ (see Bignami et al. 1996).
On the other hand, ROSAT data collected in 1993 (Halpern \& Wang 1997) yield an higher 
best fitting temperature ($T = 5.6 ~ 10^{5} ~K$) for a somewhat lower bolometric flux. 
This implies, for the same Geminga distance,  a smaller emitting area. However, the large errors on the spectral fit parameters make it difficult to draw firm conclusions on the effective size of the emitting area which could be anywhere from 10\% to the totality of the neutron star surface.

\begin{figure}
\centerline{\hbox{\psfig{figure=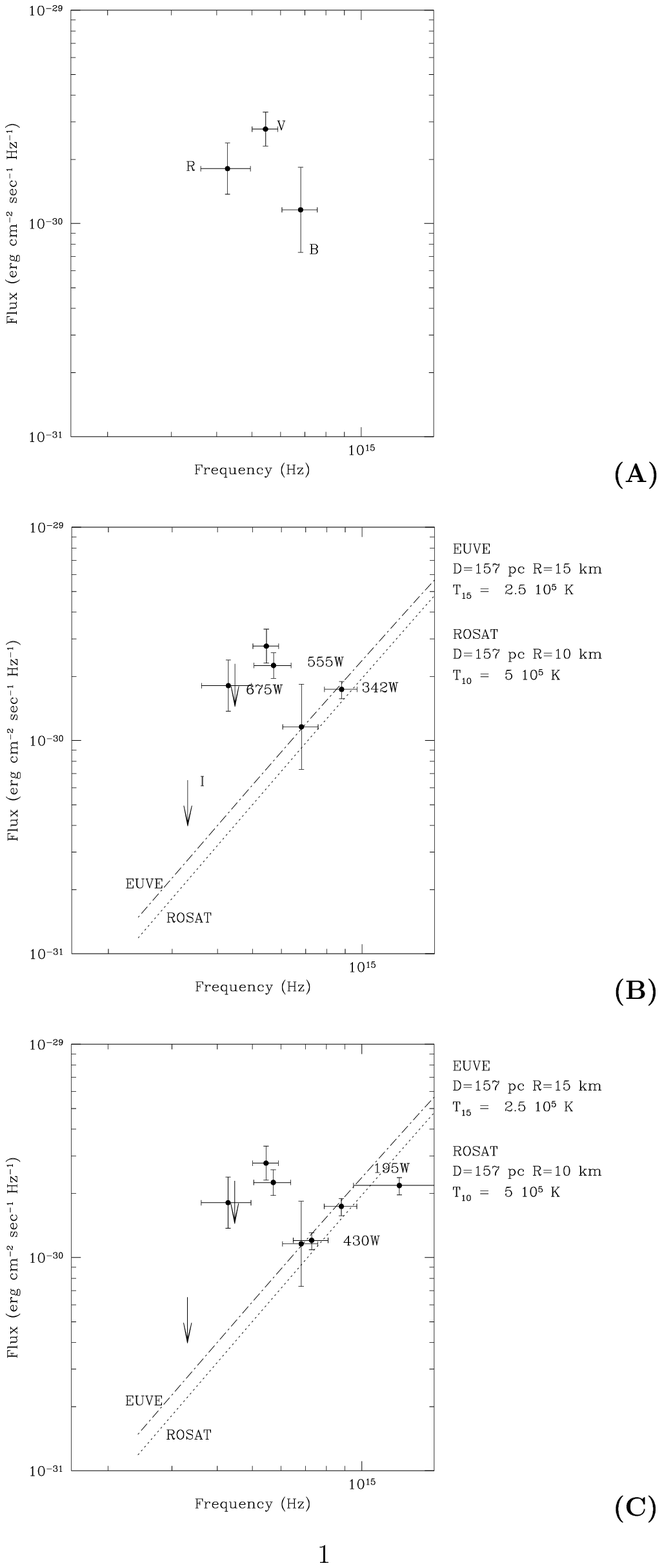,height=18cm,clip=}}}
\caption{Ten-year evolution of the  I-to-UV photometry of  Geminga.  (top) Situation in
1987, with 3 ground-based (CFHT, ESO 3.6m) points  (R,V,B)   clearly not compatible with a
black-body curve (Bignami et al. 1988). (middle) By the end of 1995,  several  points
were added (see Bignami et  al. 1996 where, indeed, a numerical error of  a factor 4 is
present in Figs. 2 and 3, where all the black-body fits should be revised downwards) both from the ground (I) and from HST (555W,
675W, 342W). (bottom) New HST/FOC data (430W, 195W) presented here.  The lines shown
represent best fit backbody curves to the  ROSAT/EUVE data for an  INS at d=157 pc
(Caraveo  et al. 1996).  The two cases shown correspond to R=10 km and  $T =4.5 \times
10^{5} $K  (ROSAT 1991 fit-dotted) and to R=15 km and T=$2.5 \times 10^{5} K$   (EUVE
fit-dashed). Note the absolute scale: no  normalization has been  performed.}
\end{figure}

\section{New data.}

Additional observations of Geminga have been obtained more recently with the HST/FOC. 
In order to have a firm confirmation of our ground-based B measure,  one observation
was performed through the 430W filter ($\lambda=3940 \AA$; $\Delta \lambda= 832 \AA$)
i.e. very close to the B band.  To further probe the source spectral shape, the second observation  was performed through the 195W one, a wide passband UV
filter ($\Delta \lambda= 946 \AA$) centered at $\lambda=2110 \AA$.   The data were
collected on May 4, 1996 with total exposure times of 8200 s with filter  430W and 5100
s with filter 195W. Following the usual strategy,  the observations were split in
shorter exposures to allow for cosmic ray removal.  The camera was operated in the
nominal "F/96" mode,  to which corresponds an effective focal ratio of f/151 and a
field of view of  $7 \times 7$ arcsec. The pixel size was thus 0.014 arcsec. \\ After
pipeline processing, single exposures were combined and  cosmic ray hits rejected by
frame to frame comparisons. Average images were thus computed  for each filter and
smoothed over cells of $3 \times 3$ pixels using a gaussian function.  In both filters,
a single point source was clearly detected in the FOC field of view.  Its coordinates coincide
with the expected position of Geminga after  correcting for the object's proper motion
(Caraveo et al. 1996) at the observing epoch. Standard FOC photometry  yielded 
magnitude values of $m_{430W} = 25.7 \pm 0.1$ and  $m_{195W} = 23.9 \pm 0.1$. 
Conversions from magnitudes to monochromatic fluxes have been computed  using the zero
points provided by the HST pipeline processing. The resulting fluxes are shown in Fig.
1c, labelled with the corresponding filter names. 

\begin{table*}
\begin{center}
\begin{tabular}{|l|l|lll|cc|}  \hline
{\em Date} & {\em Telescope} & {\em Filter} & $\lambda_{0} (\AA)$ & $\Delta \lambda (\AA)$ & {\em mag} & {\em Flux } \\
 & & & & & & $(10^{-30} ~erg/cm^{2} s ~ Hz)$ \\ \hline
Jan 94 	                  & NTT/SUSI    & I$^{2}$    &  9000 & 2400   & $\ge 26.4 $     & $\le 0.64$ \\
Jan 87 		          & CFHT    & R$^{1}$    &  7000 & 2200   & 25.5 $\pm$ 0.3  & 1.8 $\pm$ 0.5 \\
Sep95                     & HST/WFPC2   & 675W$^{2}$ &  6735 & 889.4  & $\ge$ 26        & $\le 2.3$ \\
Jan 87, Nov 92, Jan. 94   & 3.6m/EFOSC-NTT/SUSI & V$^{1,2}$  &  5550 & 890    & 25.4 $\pm$ 0.2  & 2.77 $\pm$ 0.5 \\
Mar/Sept 94, Mar 95       & HST/WFPC2   & 555W$^{2}$ &  5252 & 1222.5 & 25.5 $\pm$ 0.15 & 2.25 $\pm$ 0.3 \\
Jan 87 		          & 3.6m/EFOSC    & B$^{1}$    &  4400 & 980    & 26.5 $\pm$ 0.5  & 1.16 $\pm$ 0.6 \\
May 96 		          & HST/FOC     & 430W$^{3}$ &  3940 & 832    & 25.7 $\pm$ 0.1  & 1.20 $\pm$ 0.1 \\
Sep 94			  & HST/FOC     & 342W$^{2}$ &  3410 & 702    & 24.9 $\pm$ 0.1  & 1.73 $\pm$ 0.15 \\
May 96		          & HST/FOC     & 195W$^{3}$ &  2110 & 946    & 23.9 $\pm$ 0.1  & 2.18 $\pm$ 0.2 \\ \hline
\end{tabular}
\end{center}

\vspace{0.2cm}
$^{1}$ Bignami et al. 1988; 
$^{2}$ Bignami et al. 1996; 
$^{3}$ this work
\vspace{0.2cm}

\caption{Summary of the I-to-UV photometry of Geminga  presented in Fig. 1. Tha data have been collected over a time
span of 10 yrs both from the ground (CFHT, ESO 3.6m and NTT) and from the HST.  The columns list the observing dates,
the telescope/detector combinations, filter names, peak wavelengths and peak widths,
respectively. Three digits identify HST filters. The last two columns list the magnitude values and the corresponding fluxes, computed at the filter peak frequencies.   
For the V/555W filters, the quoted magnitudes
correspond  to the average of three, independent, exposures performed at different
epochs, aimed at measuring the proper motion (Bignami et al. 1993;
Mignani et al. 1994) and the parallax (Caraveo et al. 1996). The only
other optical data on Geminga are the early ones of Halpern \& Tytler (1988) taken
through Gunn g and r filters and largely consistent with those reported above.}
\label{}
\end{table*}

\section{Interpretation.}

It is apparent  that the new, more accurate,
FOC 430W point confirms the B-flux obtained back in 1988 at ESO's 3.6m. At the same
time, the 195W point is certainly higher than the 430W and also of the 342W one. 
Moreover,  the V flux value was independently confirmed by  repeated observations 
both from the ground (NTT) and from the HST (Caraveo et al. 1996).
As shown in Fig.1, our set of optical observations is self-consistent, repeatable, and robust. 
It must, therefore, represent a firm anchor for the spectrum of Geminga. \\
The points blueward of the 
430W have a slope relative to each other certainly consistent with the Rayleigh Jeans $E^{2}$ slope. This is why it appears natural to regard them as the optical-UV emission from an X-ray planckian, provided, of course, that its temperature be consistent with the observed flux. This was the scenario arising from the ROSAT data published in Halpern \& Ruderman (1993), where everything, i.e. X-ray temperature, neutron star dimension and distance seemed to fall in place on the optical-UV points given above. In such a scenario, Occam's razor rendered unescapable an interpretation centered on single optical-to-$\sim$ 1 keV  thermal emission  from a 
$\sim 500,000~K$ neutron star at the Geminga distance. On top of this, the V-centered feature was interpreted as a wide cyclotron line, for want of any other possibility. \\ Such a simple scenario may now have to be changed if one is to accept the more recent data of Halpern \& Wang (1997), who both increase the best fitting X-ray temperature and decrease the total flux and thus the fraction of the emitting surface. Accepting this new X-ray interpretation implies that the optical data can no longer be considered as part of the same thermal emission generating the X-rays. However, we would like to underline the robustness of our repeated optical-UV data which feature error bars very much smaller than those characterizing any X-ray extrapolation.\\ Globally, considering the big uncertainties of the X-ray-derived parameters (to wit Halpern \& Ruderman, 1993, vs. Halpern \& Wang, 1997), we would still be inclined towards the simplest interpretation, i.e.  to ascribe this feature to an emission line,
superimposed to the $E^{2}$ Rayleigh-Jeans continuum.
A fitting procedure assuming a simple Gaussian line centered at $\lambda_{c}$, with 
FWHM=$\Delta \lambda_{c}$, and a total energy flux  $F_{c}$, yields a very good fit
($\chi^{2}/d.o.f. = 0.8$) to the data with the following  parameters (at infinity):
$\lambda_{c}= 5998  \AA$,  $\Delta \lambda_{c} \sim 1,300 \AA$,  
$F_{c} \sim 4 \times 10^{-16} erg~cm^{-2}s^{-1}$. \\ 
As mentioned in Bignami et al. (1996), the interpretation of such a feature is based on the
presence of an atmosphere on  Geminga,  something already  imagined for INSs in general, as
well as explicitly for Geminga (Meyer
et al 1994). For the  case of an atmospheric cyclotron-emitting
layer the observed $\lambda_{c}=5998 \AA$  (= 2.06 eV) implies a B field, measured 
at infinity, of $\sim 3.8~
10^{11} G$ for Z/A = 1, i.e. pure H atmosphere, and correspondingly higher in the presence of
He or heavier elements.  \\ This would be the first direct magnetic field measurement
for an INS.  Of course, for neutron
stars  in binary systems, direct measures of the magnetic fields were reported by Tr\"umper et al. (1978). In particular, our value
compares  favourably with the  theoretical one $B \sim (P \dot P)^{1/2}$ , which for
the  Geminga parameters  (Mattox et al. 1998) turns out to be $\sim 1.5 \times 
10^{12}~G$.    The apparent huge width  of the feature ($\sim 1,300 \AA$), or  20\%
considering its centre at  $\sim 6,000 \AA$, could  be accounted for by the spread of
the B-field values integrated in space and time over the whole rotating NS surface. \\
Alternative views, including the possibility of a significanty non-thermal component, should certaily be considered. This will be particularly true if the recent exciting result on the detection of an optical pulsation (Shearer et al. 1997a) is confirmed and the shape of the optical light curve becomes available for comparison, especially with the X-ray data. 

\section{Conclusions.} 

The new data presented here, improving as they do the data given in Bignami  et al.
(1996), leave no room for doubt
that a wide emission feature exists around  $6,000 \AA$, superimposed on a possibly thermal
continuum.  The feature falls in the wavelength region  expected for an atmospheric
proton-cyclotron emission produced close to the surface of a magnetic neutron star.
Geminga is a magnetic neutron star, to wit its periodic $\gamma$-ray emission
(Mattox et al, 1998), and most probably has an atmosphere, to wit the spectral shape
of  its soft X-ray emission (Meyer et al 1994, Halpern \& Rudermann 1993). \\ Not only the
optical/UV data may represent, historically, the first time we  actually  see  the surface 
(atmosphere) of an INS: our interpretation of the feature reported here
would  also imply the first direct measure of the magnetic  field of an INS.\\ The
question arises wheter to expect similar features for other INSs, as the
X-ray/UV/optical data improve. If the physical assumptions mentioned above are correct,
then, by and large, the answer is positive. We do expect to observe similar features on
other INSs of comparable age, first of all on the other two "musketeers".  However, for
PSR0656+14 the photometry information available  so far
(Pavlov et al. 1997) shows a rather different spectral behaviour and the
optical flux  appears  more consistent with a steep power law ($\alpha \sim 1.4$),
although an additional blackbody component is present.  Its optical emission would
thus be the combination of both magnetospheric and thermal processes.  The recent
detection of optical pulsations from PSR0656+14 (Shearer et al. 1997b) seems also to
support the presence of a strong  magnetospheric component. For the third and oldest
musketeer, our recent detection (Mignani et al. 1997) falls close enough to the
Rayleigh-Jeans extrapolation of the ROSAT spectrum. \\
The difficulties in understanding INS physical properties have been outlined dramatically, 
even for the three best known cases of X-ray thermal emitters. Of them, Geminga is certainly the one with the most abundant phenomenology at yet, disentangling its thermal and non-thermal emission (i.e. distiguishing surface from magnetosphere), although well under way, will require more refined observations. 

\begin{acknowledgements}
R. Mignani is grateful to Joachim Tr\"umper for the critical reading of the manuscript.      
\end{acknowledgements}

\end{document}